\documentclass[superscriptaddress, preprint, aps]{revtex4-2}
\usepackage{graphicx}
\graphicspath{ {../fig/}}

\usepackage{amssymb}
\usepackage{amsmath}
\usepackage{amsthm}
\usepackage{amsfonts}
\usepackage{xfrac}
\usepackage{wasysym}
\usepackage{textcomp}
\usepackage{bbm}

\usepackage{braket}
\usepackage[version = 4]{mhchem}
\usepackage{epstopdf} 

\usepackage{hyperref}
\usepackage{array}
\newcommand{\figSize}{0.95}

\begin{document}

\author{Stefano Marin}
\email{stmarin@umich.edu}
\affiliation{Department of Nuclear Engineering and Radiological Sciences, University of Michigan, Ann Arbor, MI 48109, USA}

\author{Eoin P. Sansevero}
\email{eoins@umich.edu}
\affiliation{Department of Nuclear Engineering and Radiological Sciences, University of Michigan, Ann Arbor, MI 48109, USA}

\author{M. Stephan Okar}
\affiliation{Department of Nuclear Engineering and Radiological Sciences, University of Michigan, Ann Arbor, MI 48109, USA}

\author{Isabel E. Hernandez}
\affiliation{Department of Nuclear Engineering and Radiological Sciences, University of Michigan, Ann Arbor, MI 48109, USA}

\author{Ramona Vogt}
\affiliation{Nuclear and Chemical Sciences Division, Lawrence Livermore National Laboratory, Livermore, CA 94550, USA}
\affiliation{Physics and Astronomy Department, University of California, Davis, CA 95616, USA}

\author{J\o rgen Randrup}
\affiliation{Nuclear Science Division, Lawrence Berkeley National Laboratory, Berkeley, California 94720, USA}



\author{Shaun D. Clarke}
\affiliation{Department of Nuclear Engineering and Radiological Sciences, University of Michigan, Ann Arbor, MI 48109, USA}

\author{Vladimir A. Protopopescu}
\affiliation{Oak Ridge National Laboratory, Oak Ridge, TN 37830, USA}

\author{Sara A. Pozzi}
\affiliation{Department of Nuclear Engineering and Radiological Sciences, University of Michigan, Ann Arbor, MI 48109, USA}
\affiliation{Department of Physics, University of Michigan, Ann Arbor, MI 48109, USA}

\title{Directional-dependence of the event-by-event neutron-$\gamma$ multiplicity correlations in $^{252}$Cf(sf)}
\date{\today}

\begin{abstract}
 We differentiate the event-by-event n-$\gamma$ multiplicity data from \ce{^{252}Cf}(sf) with respect to the energies of the emitted particles as well as their relative angles of emission. We determine that neutron emission enhances $\gamma$-ray emission around $0.7$ and $1.2$ MeV, but the only directional alignment was observed for $E_\gamma \leq 0.7$ MeV and tended to be parallel and antiparallel to neutrons emitted in the same event. The emission of $\gamma$ rays at other energies was determined to be nearly isotropic. The presence of the emission and alignment enhancements is explained by positive correlations between neutron emission and quadrupole $\gamma$-ray emission along rotational bands in the de-exciting fragments. This observation corroborates the hypothesis of positive correlations between the angular momentum of a fragment and its intrinsic excitation energy. The results of this work are especially relevant in view of the recent theoretical and experimental interest in the generation of angular momentum in fission. Specifically, we have determined an alignment of the fragments angular momenta in a direction perpendicular to the direction of motion. We interpret the lack of $n$-$\gamma$ angular correlations for fission fragments near closed shells as a weakening of the alignment process for spherical nuclei. Lastly, we have observed that statistical $\gamma$ rays are emitted isotropically, indicating tha the average angular momentum removed by this radiation is small. These results, and the analysis tools presented in this work, representing a stepping stone for future analysis of $n$-$\gamma$ emission correlations and their connection to angular momentum properties. 
\end{abstract}


\maketitle

\section{Introduction} 
\label{sec:intro}

Neutrons and $\gamma$ rays emitted from fission fragments reveal important features of the nuclear fission process and the state of the fragments immediately following fission. Among several open questions in fission, the $n$-$\gamma$ angular correlations are particularly interesting because of their intimate relation to the fission fragment angular momenta. The angular momentum of a fragments plays a pivotal role in the emission of $\gamma$ rays and the $n$-$\gamma$ angular distribution. The characterization of the fragment angular momenta is one of the most important open questions in fission physics.

The first experimental investigations of fission fragment angular momenta were carried out in the 1960s~\cite{Hoffman1964} and 1970s~\cite{Wilhelmy1972} and
some early theoretical work was done in the 1980s on the character 
of the distribution of the fragment angular momentum \cite{Moretto1980} and
the underlying mechamism for its generation \cite{Dossing1985}. 
Due to the advances in instrumentation, modeling, and computation since then, the topic of fission fragment angular momenta has gained renewed interest in recent years~\cite{Stetcu2021, bulgac2021fragment, Vogt2021, Randrup2021, Wilson2021, Chebboubi2017, Rakapoulos2018, Marevic2021}. Of particular interest are the event-by-event correlations between fragment energy and angular momentum, the directional alignment of the angular momentum with respect to the motion of the fragment, and the correlations, both in magnitude and direction, between the angular momenta of the two fragments.

The evaporation of neutrons from fragments is highly correlated with the fragment intrinsic excitation energy~\cite{Gook2014, Signarbieux1972, Gavron1971}, whereas $\gamma$-ray emission correlates strongly with the fragment angular momentum~\cite{Travar2021, Nifenecker1972}. By analysing the correlations between neutrons and $\gamma$ rays it is possible to infer the underlying correlations between the fragment energy and angular momentum. To reduce noise and systematic biases associated with the emission of these particles, it is necessary to differentiate the emission based on the kinematic properties of the emission, namely their kinetic energies and directions. 

In this work, we continue our analysis of the event-by-event neutron-$\gamma$ multiplicity correlations presented in Refs.~\cite{Marin2020, Marin2021}. In Ref.~\cite{Marin2020}, we determined that neutron and $\gamma$-ray emissions are slightly negatively correlated, as a result of energy and angular momentum conservation. The more recent analysis of Ref.~\cite{Marin2021} analyzed how the correlations depend on the energy of the emitted particles. We have observed predominantly negative correlations with notable positive enhancements at specific $\gamma$-ray energies: $E_\gamma \approx 0.7$ and $1.2$ MeV. With the aid of model calculations, we concluded that the positive enhancements originated from positive correlations between the angular momenta and energies of the fragments in a fission event. In this work, we extend the previous investigations by analysing the correlations differentiated with respect to both energy and direction. 
The paper is structured as follows. In Section~\ref{sec:sources}, we discuss the origins of the angular distribution of neutron and $\gamma$ radiation. In Section~\ref{sec:analysis}, we present the new analysis of the experimental data that takes into account both the energy and the angular dependence of the emitted radiation. Section~\ref{sec:results} presents the analysis of the data collected using the Chi-Nu liquid organic scintillator array at LANSCE, Los Alamos. The experimental results and possible theoretical interpretations are also discussed. Lastly, in Section~\ref{sec:conc} we discuss how the observed $n$-$\gamma$ emission alignments, and lack thereof, indicate that the fragment angular momentum is polarized in a direction perpendicular to fragment direction of motion, and the relationship between the fragment angular momentum and excitation energy.

\section{Sources of Angular Correlations} 
\label{sec:sources}

In the fragment center-of-mass frame (CoM), neutrons are emitted with mean velocities comparable to the speed of the fragment in the lab frame. Thus, the neutron kinematic boost effectively determines the angular distribution of neutrons in the lab frame. The emission of neutrons in the CoM is often approximated as isotropic and this approximation has been validated experimentally \cite{Chietera2015, Gook2014}.  Other effects, related to the coupling of angular momenta and the possibility of scission neutrons~\cite{Petrov2008}, would only result in small corrections. Thus, we expect that neutrons will primarily follow the direction of the fragment motion. Because the light and heavy fragments are emitted emitted back-to-back in the CoM of the initial fissioning nucleus, the angular distribution of neutrons appears as two distributions focused parallel and anti-parallel to the motion of the fragments, \textit{i.e.}, the fission axis. 

The kinematic focusing causes neutrons with greater kinetic energy in the lab frame to be more tightly aligned with the fission axis and their distributions more anisotropic in the lab frame. Larger lab-frame energies are also associated with larger CoM energies, which biases the sample toward symmetric fission (see Figs. 6 and 18 in Ref.~\cite{Gook2014}), \textit{i.e.}, fission events resulting in two similar mass fragments.

The angular distribution of $\gamma$ rays is also affected by kinematic boosting, an effect known as $\gamma$-ray aberration~\cite{landauEM}. However, the effects are significantly weaker given the relatively low velocity of fragments. The effects of a weak aberration depend on the angular distribution in the CoM, but can be approximated as a linear term in the cosine of the angle of emission in the lab frame. The kinematic boosting of both neutrons and $\gamma$ rays tend to make $n$-$\gamma$ angular distribution more parallel when the particles are emitted by the same fragment, and more antiparallel when the particles are emitted by different fragments. Because we observe $n$-$\gamma$ correlations irrespective of the fragments emitting them, these two effects tend to cancel each other. Thus, we do not expect the aberration of $\gamma$ rays to have a dominant role in $n$-$\gamma$ angular correlations. 

The coupling of the fragment angular momentum with that of emitted $\gamma$ rays gives rise to strong observable angular correlations~\cite{Hoffman1964}. We can observe the angular correlations of $\gamma$ rays relative to the fission axis because the angular momenta of the fragments is aligned perpendicular to the fission axis~\cite{Hoffman1964, Randrup1982, Moretto1980, Randrup2014, Randrup2021}. 


The emission of $\gamma$ rays following fission is usually divided into two stages: first, $\gamma$ rays are emitted in the continuum to dissipate the intrinsic excitation energy left over after neutron evaporation; second, $\gamma$ rays are emitted to dissipate the energy stored in the collective degrees of freedom. We call the first type of $\gamma$ emission \textit{statistical}, because the transition strengths are determined from a statistical analysis of the level densities. We call the second type of emission \textit{discrete}, since the transitions are determined by the available levels in the discrete region of the level scheme.  The angular distributions of these two categories of $\gamma$ rays are very different.

Statistical $\gamma$-ray emission is assumed to be primarily electric dipole radiation. The angular distributions of statistical $\gamma$ rays has been described to be either isotropic~\cite{Hoffman1964}, or aligned parallel to the fragment angular momentum and thus perpendicular to the fission axis~\cite{Oberstedt2018}. The difference between these two alternatives lies in the angular momenta of the initial and final states, $J_i$ and $J_f$ respectively.  Transitions with $J_f = J_i \pm 1$ contribute $\gamma$ rays emitted predominantly perpendicular to the fission axis, whereas $J_f = J_i$ contribute $\gamma$ rays emitted predominantly parallel to it~\cite{Tolhoek1953}. Depending on the proportion of the two types of dipole transitions, the angular distributions of statistical $\gamma$ rays can have different angular distributions. 

Discrete emission along the yrast band is primarily electric quadrupole in nature, although magnetic dipole contributions at the lowest energies have also been observed. Discrete quadrupole emission along a rotational band tends to be stretched, i.e., the angular momentum removed by the radiation is maximized, $J_f = J_i -2$. Because of their stretched character, the angular distribution of $\gamma$ rays from quadrupole band transitions are directed approximately perpendicular to the angular momentum axis and are thus predominantly parallel to the fission axis~\cite{Tolhoek1953}. 


Based on the discussion presented above, we expect both neutron and $\gamma$-ray emission to be correlated with the fission axis, with no directly correlations between them. Intrinsic correlations between sequential emissions are possible, even in the case of a nucleus that is not initially oriented. These angular correlations arise because the fragment, as it de-excites from energy level to energy level, is in a superposition of magnetic substates of angular momentum. Angular momentum conservation dictates that the magnetic quantum numbers of successive levels are entangled with one another, introducing intrinsic correlations between them.

The intrinsic angular correlations between neutrons and $\gamma$ rays can be quite strong. Thus, while we cannot currently exclude that these angular correlations play an important role in the determination of $n$-$\gamma$, there are several factors that reduce their strength. First, we expect these intrinsic correlations only to affect the emission from a single fragment. Second, the emission of other particles in the same decay sequence will diminish the observed correlations. This is particularly important for the correlations between neutrons and the discrete $\gamma$ rays, generally emitted after $1-2$ statistical $\gamma$ rays have been emitted. Third, intrinsic correlations can only be expected if neutrons are emitted with some orbital angular momentum. It is usually assumed that neutrons below approximately $1-2$ MeV are emitted predominantly as $s$-waves, thus significantly reducing the effects of intrinsic angular correlations. A recent investigation~\cite{Stetcu2021} showed that the optical model of the nucleus can predict much larger values of the neutron orbital angular momentum, but more evidence is needed. In light of these considerations, the intrinsic angular correlations are not explicitly discussed in this paper, and will be investigated in future work. 





\section{Experimental Analysis} 
\label{sec:analysis}

 In this analysis, we differentiate the covariance of the event-by-event neutron and $\gamma$-ray multiplicities, $N_n$ and $N_\gamma$, with respect to the angle between them, $\theta_{n \gamma}$, as well as their energies, $E_n$ and $E_\gamma$. This analysis is an improvement of the  analysis presented in  Ref.~\cite{Marin2021}, which only differentiated correlations with respect to energy.
 The differentiated normalized covariance,  $C_{E_n E_\gamma \theta_{n \gamma}}$, is 

\begin{equation}
    C_{E_n E_\gamma \theta_{n \gamma} } = 
    \frac{\partial^3 \text{cov}(N_n, N_\gamma)}{\partial E_n \partial E_\gamma \partial \theta_{n \gamma} } 
    \left[
    \frac{\partial}{\partial \theta_{n\gamma} }
    \left(
    \frac{\partial \langle N_n \rangle }{\partial E_n}
    \frac{\partial \langle N_\gamma \rangle }{\partial E_\gamma}
    \right)
    \right]^{-1} \ .
\label{eq:diffCov}
\end{equation}
The quantity $C_{E_n E_\gamma \theta_{n \gamma}}$ is bounded from below at $-1$, but has no upper bound. The three differentiations we perform here serve distinct purposes. Because the discrete level transitions tend to be lower in energy than statistical emission, the differentiation with respect to $E_\gamma$ helps to sharpen the separation between statistical and discrete emission. The differentiation with respect to neutron energy sharpens the neutron-$\gamma$ angular correlations by narrowing the angular distribution of neutrons with respect to the fission axis. This differentiation also allows the identification of correlations that exist due to sample biasing, and are thus unrelated to the more interesting correlations between a fragment energy and angular momentum. Lastly, the differentiation with respect to angle is used to identify the  angular momentum properties of $\gamma$ rays. 



The data analyzed in this paper were collected with the Chi-Nu array at Los Alamos National Laboratory. These data are the same as those analyzed in Refs.~\cite{Marin2020, Marin2021, Marcath2018, Schuster2019}. A detailed description of the experiment and the detector can be found in those references. In this paper, we focus on the capabilities of angular measurements with the Chi-Nu array. See Ref.~\cite{Marin2021} for a discussion of the energy acceptance of the Chi-Nu detectors. 

Because spontaneous fission lacks a preferred direction and the fission axis is not experimentally measured, the directional distribution is measured between pairs of emitted particles. Specifically, in the present experiment we measure the neutron-$\gamma$ covariance between the measured neutron and $\gamma$-ray multiplicities in two detectors whose geometric centers are separated by an angle $\theta_{n \gamma}$. The normalization of the covariance, the factor in square brackets in Eq.~\eqref{eq:diffCov}, is calculated from the product of the mean measured multiplicities in each detector. Because of different gain settings and distances to the source, the efficiencies of each detector varied considerably. These variations in the individual detector behavior are directly translated into variations in the product of efficiencies for each detector pair. With $42$ active detectors during the experiment, a total of $861$ detector pairs are possible. However, because we take the first detector to detect a neutron and the second a $\gamma$-ray, each detector pair is doubly degenerate. Thus, a total of $1722$ detector-pair combinations are considered. 

\begin{figure}[!htb]
\centering
\includegraphics[width=\figSize \linewidth]{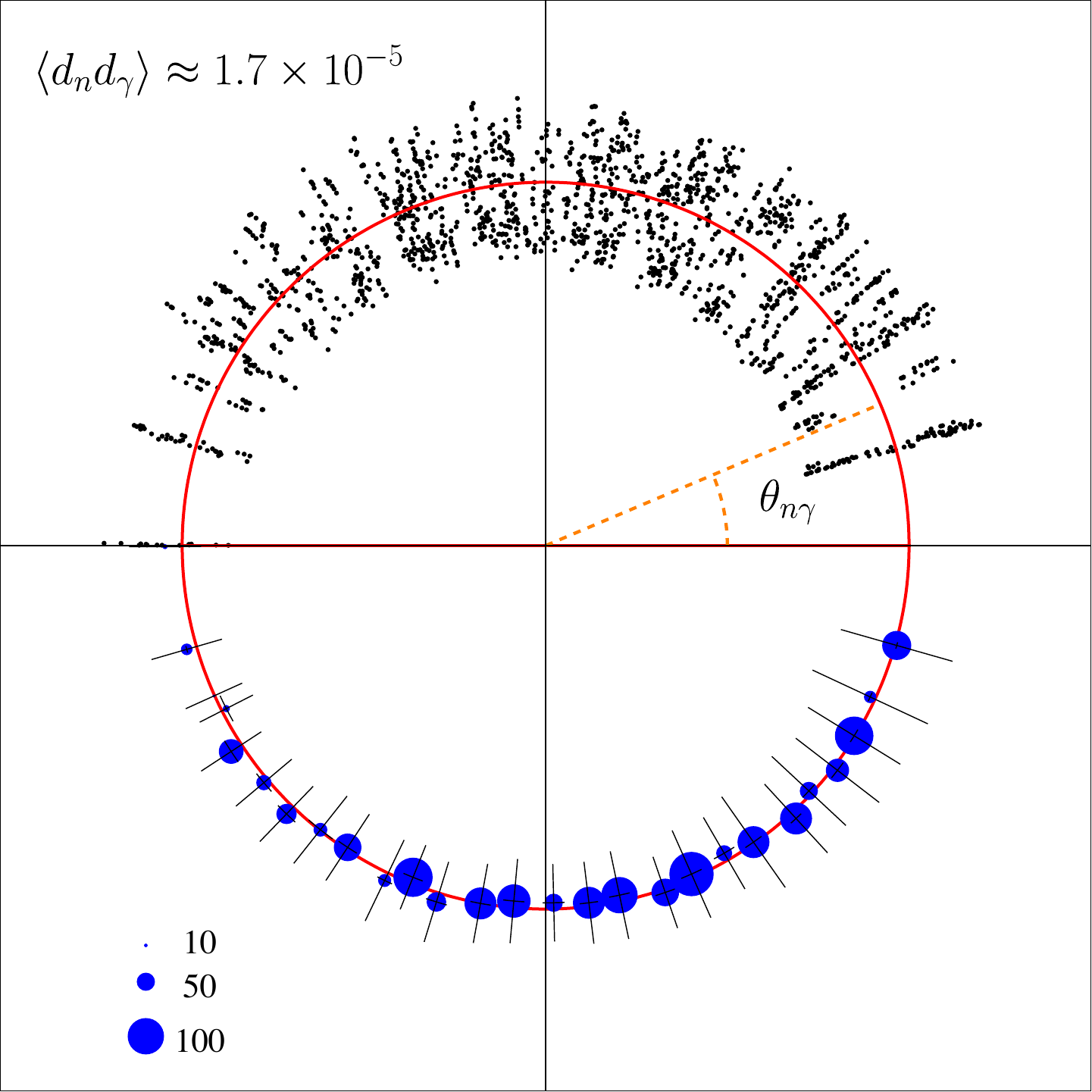}
\caption{The angular efficiency of the Chi-Nu detection system. See text for details. } 
\label{fig:effChiNu}
\end{figure}

We show the angular efficiency of the system in the upper half of Fig.~\ref{fig:effChiNu}. Each point in the polar plot represents a detector pair while the distance from the origin represents the energy-averaged efficiency of the detector pair, the product in the measured neutron-$\gamma$ multiplicities across all energies. The red circle represents the average detector-pair efficiency across all detector pairs. The average detector efficiency, expressed as the rate of double counts per fission event in a specific detector pair $\langle d_n d_\gamma \rangle$, is indicated on the figure. 
In the lower half of Fig.~\ref{fig:effChiNu}, we group detector pairs in angular bins. The size of the marker represents the number of detector pairs included in that group, while the position of the marker represents the average in angle and efficiency for all pairs in the group. The legend indicates the number of detector pairs in each bin. The standard deviations in both angle and efficiency are shown as error bars. 

Because the Chi-Nu array is hemispherical, angles between detector pairs are neither isotropic nor symmetric with respect to $\pi/2$. The variations observed in the efficiencies are significantly reduced if the mean of each angular group is taken. In fact, we see from the lower half of Fig.~\ref{fig:effChiNu} that the mean pair efficiency of the angular groups falls close to the average over all detector pairs. However, it should be noted that because of the hemispherical geometry of Chi-Nu, we expect measurements to be biased toward neutrons and $\gamma$ rays emitted at acute angles.

The angular resolution of the detection system is defined as the spread in the angles measured by the experimental system when the particles are emitted at a fixed angle. The resolution depends on the room return, where the radiation interacts with materials around the detector system, or the detector system itself, before being measured, and on the finite width of the detectors. Given the large size of the liquid organic scintillators employed, the resolution due to finite width of the detectors dominates, introducing an uncertainty in the measured angles of $\approx 0.13$ rad. This result was confirmed by an MCNPX-PoliMi~\cite{Pozzi2003} simulation of the detector array. Using the same simulation, we have also determined that systematic biases are negligible: the mean angles between measured particles will be the same as the emitted angles, while the width of the measured angular distribution is broadened by angular resolution effects. Cross talk effects are negligible because neutrons and $\gamma$ rays are easily distinguished by the organic scintillators. 

The unfolding of the neutron and $\gamma$-ray energies is performed using the method described in Ref.~\cite{Marin2021}. This type of unfolding does not completely recover the initial distribution, but addresses systematic biases in the average spectra. Specifically, the unfolded distribution remains broadened with a system-characteristic resolution. Because the angular response of Chi-Nu does not introduce systematic biases and only introduces broadening, we do not unfold the angle and take the angle between two particles to be the angle between the centers of the detectors where the particles interacted. We consider energy ranges of  $1.0 < E_n < 7.4$ MeV and $0.24 < E_\gamma < 3$ MeV, with bin widths of size $0.4$ and $0.16$ MeV for neutrons and $\gamma$ rays, respectively. 

\section{Results} 
\label{sec:results}


 For every fixed neutron and $\gamma$-ray energy $E_n$ and $E_\gamma$, we obtain curves defining the magnitude of neutron-$\gamma$ correlations for angles $\theta_{n\gamma}$ between the two emissions and extract the Legendre coefficients from them. In the CoM frame, only even-order polynomials can describe the angular distribution of $\gamma$ rays~\cite{Tolhoek1953}. However, the aberration of $\gamma$ rays can introduce odd-order, antisymmetric terms to the angular distribution. 
 
 We have fit the experimental data using Legendre polynomial, and have determined that the best fit was provided by retaining only the symmetric 0$^{\rm th}$ and 2$^{\rm nd}$ order polynomials. Thus, the $n$-$\gamma$ correlations at fixed energies are fit using

\begin{equation}
    C_{E_n E_\gamma \theta_{n\gamma}} = A_0(E_n, E_\gamma) + A_2(E_n, E_\gamma) P_2 (\text{cos}\theta_{n\gamma} ) \ .
    \label{eq:legCoeff}
\end{equation}
%

\begin{figure}[!htb]
\centering
\includegraphics[width=0.8\linewidth]{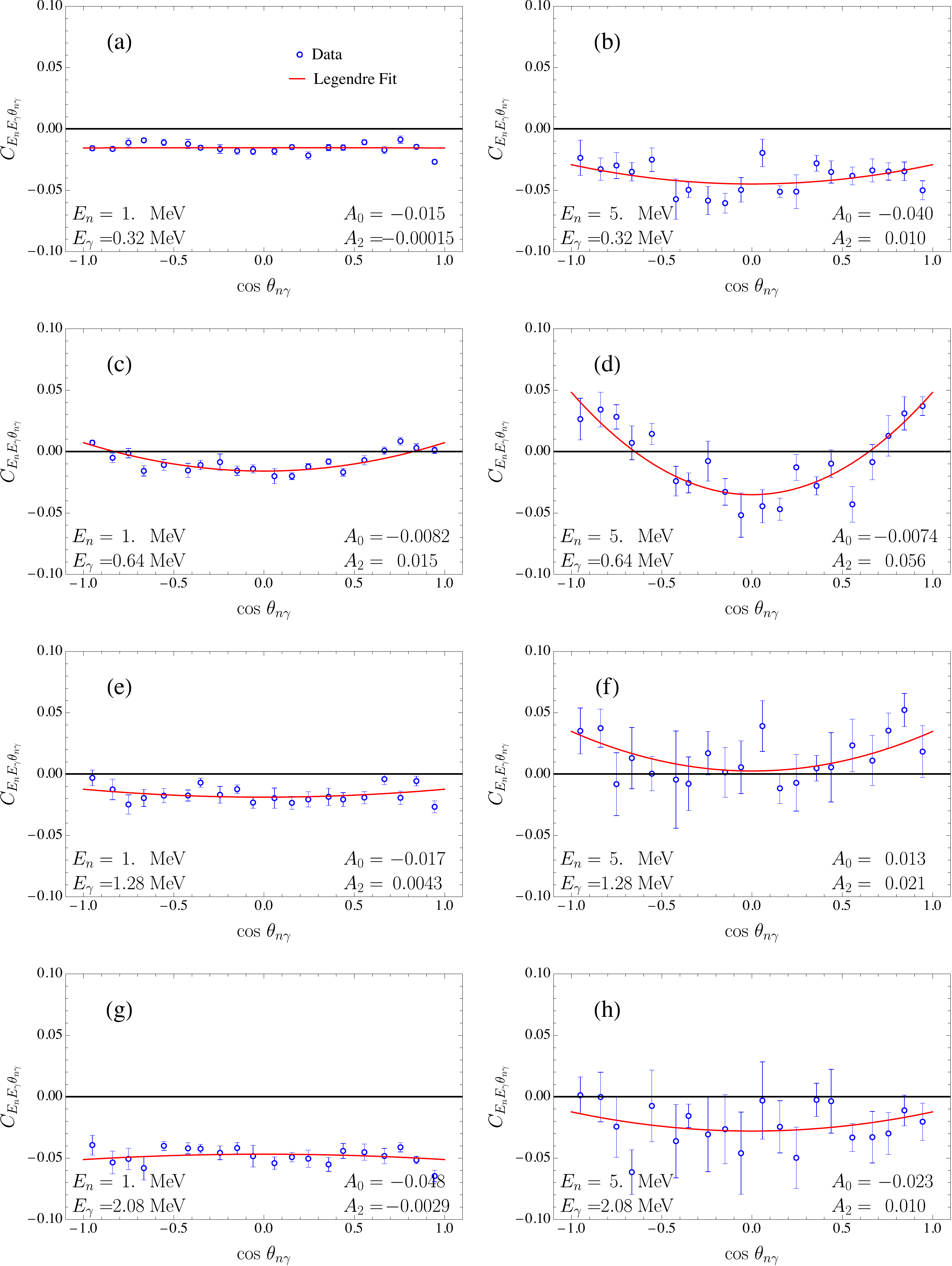}
\caption{Curves of $C_{E_n, E_\gamma \theta_{n \gamma}}$ for selected neutron and $\gamma$-ray energies. Legendre polynomial fits for $0^{\text{th}}$ and $2^{\text{nd}}$ order are shown.} 
\label{fig:demoFit}
\end{figure}

\begin{figure}[!htb]
\centering
\includegraphics[width=\figSize \linewidth]{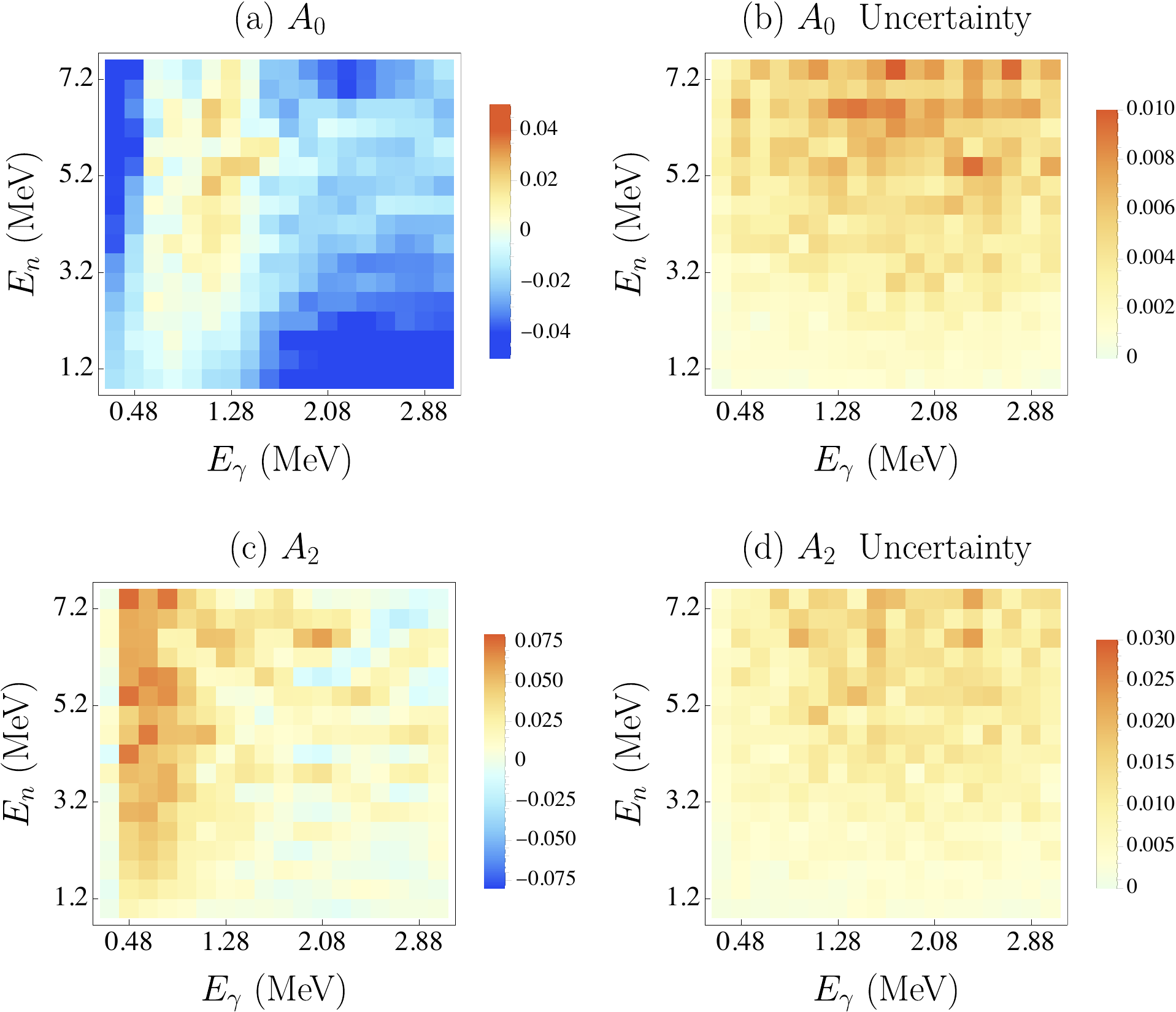}
\caption{Legendre polynomials fit parameters to $C_{E_n E_\gamma, \theta_{n \gamma}}$} 
\label{fig:legCoeff}
\end{figure}

Neutron-$\gamma$ correlation obtained from the data are shown in Fig.~\ref{fig:demoFit} for several selected combinations of $E_n$ and $E_\gamma$. On the same plot, we show the fit to the data retaining only the 0$^{\rm th}$ and 2$^{\rm nd}$ order polynomials. The Legendre coefficients across all neutron and $\gamma$-ray energies are shown in Fig.~\ref{fig:legCoeff}. Lastly, the statistical uncertainties of the coefficients, determined from randomly resampling the data are also shown in Fig.~\ref{fig:legCoeff}. 
%

The parameter $A_0$ has a simple physical interpretation: it represents the magnitude of the $n$-$\gamma$ covariance averaged over all emission angles. The result shown in Fig.~\ref{fig:legCoeff} (a), as expected from our previous investigation~\cite{Marin2021}, shows structure developing in the regions $E_\gamma \approx 0.7$ and $E_\gamma \approx 1.2$ MeV. Using model calculations, we showed that the presence of the enhancement at $0.7$ MeV can be explained by positive correlations between the fragment angular momentum and energy, increasing the feeding of rotational band states with increasing neutron multiplicities, and thus excitation energy. The enhancement at $1.2$ MeV is explained in part by the same correlations and in part by a biasing of the fission samples towards symmetric fission. For a more detailed analysis of the angle-independent $n$-$\gamma$ correlations see Ref.~\cite{Marin2021}.

The coefficient of the second Legendre polynomial, $A_2$, shown in Fig.~\ref{fig:legCoeff} (c) shows the dependence of the correlations on the emission angle between neutrons and $\gamma$ rays. Positive $A_2$ indicates $\gamma$ rays are aligned predominantly along the direction of neutron emission, both parallel and antiparallel, while negative $A_2$ indicates $\gamma$ rays are aligned perpendicular to neutron emission. 

The statistical uncertainties for both $A_0$ and $A_2$ are shown in Fig.~\ref{fig:legCoeff} (b) and (d), respectively. The uncertainties are larger for the higher energies, where fewer particles were measured. However, the uncertainties are several times smaller than the magnitude of the Legendre coefficients in the regions of enhancement that we discuss below. 


We note enhanced positive structure at $E_\gamma \approx 0.7$ MeV in $A_2$. This enhancement closely resembles the structure observed in $A_0$, but extends to lower energies and does not extend to higher energies. Importantly, we do not observe pronounced angular correlations at $E_\gamma \approx 1.2$ MeV, as we do in $A_0$. We observe a trend of enhanced correlations with increasing neutron energies. These results are not surprising considering the discussion presented in Sec.~\ref{sec:sources}. With increasing neutron lab-frame energies, we bias towards more kinematically-boosted neutrons, thus the angle between $\gamma$ rays and neutrons becomes more representative of the angle between $\gamma$ rays and the fission axis. At high $E_\gamma$ the angular correlations are much smaller and very close to $0$. Angular correlations are also weak at the lowest $\gamma$-ray energies, but caution should be used in interpreting this region as it borders the lower edge of the $E_\gamma$ acceptance and the unfolding might lead to artifacts. We do not observe significant dependence on neutron energy in either the low or high $E_\gamma$ region. 

The alignment of $\gamma$ rays with the fission axis at $E_\gamma \approx 0.7$ MeV indicates that these $\gamma$ rays are predominantly from stretched quadrupole transitions along rotational bands. As noted above, these transitions generate $\gamma$ rays predominantly perpendicular to the angular momentum, and thus parallel to the fission axis. This is not the first experimental observation of these angular correlations, see for example Refs.~\cite{Valski1969, Oberstedt2019, Hoffman1964}. However, the results of this work combine these angular correlations with the observed positive overall covariance, manifested as $A_0$,  between neutrons and $\gamma$ rays in this energy region, thus giving further confirmation of the presence of positive correlations between the fragment angular momentum and energy. 

The positively-correlated region at $E_\gamma \approx 1.2$ MeV shows some deviations from the expected behavior. In Ref.~\cite{Marin2021}, we identified high-energy transitions in spherical nuclei near the shell closure of \ce{^{132}Sn} as the main contributor to the enhanced correlations. We expect these $\gamma$ rays to be predominantly stretched, thus giving rise to positive angular correlations. However, experimental observation shows that the angular correlations in this region, while still positive, are significantly reduced with respect to the enhancement at $E_\gamma \approx 0.7$ MeV. This reduction can occur if, in spherical nuclei, the direction of the angular momentum is not strongly oriented perpendicular to the fission axis, as is the case for more deformed fragments. 

In the higher-energy region of $E_\gamma \gtrapprox 1.8 $ MeV the emission should be dominated by statistical transitions. We observe that the $\gamma$ rays in this region are predominantly isotropic. This is in good agreement with earlier observation by Val'skii \textit{et al.}~\cite{Valski1969} and Hoffman~\cite{Hoffman1964}, but in disagreement with the simplified model of stretched $E1$ transitions mentioned by Oberstedt \textit{et al.}~\cite{Oberstedt2019}. Therefore, the results of this analysis indicate that the statistical transitions in the continuum are not dominantly stretched: a significant component connects states of equal angular momentum. 
In the low-energy region of $E_\gamma \approx 0.4$ MeV we again find $\gamma$ rays uncorrelated with the neutron direction, and hence the fission axis. Val'skii~\cite{Valski1969} investigated the multipolar character of $\gamma$-ray transitions and determined that, at the lowest energies and, significantly for $E_\gamma < 0.5$ MeV, $M1$ transitions become important. We can explain the relative isotropy of these transitions at these energies, also observed by Val'skii, by considering that these intraband transitions have a lower probability of being stretched since, in some situations, two connected levels in different bands will have the same angular momentum. Even for stretched transitions, the angular momentum at low $E_\gamma$ will be reduced because the contribution of $E2$ and $M1$ transitions are both important in this energy region and their angular distributions carry opposite signs.

\section{Concluding remarks}
\label{sec:conc}

We have expanded our previous analysis of event-by-event $n$-$\gamma$ emission correlations by considering the angle between the emitted particles in addition to their individual energies. We have observed enhanced emission correlations as well as alignments of the emitted particles for $\gamma$-ray energies associated with rotational band transitions. We conclude that this enhancement is related to positive correlations between the fragment energies and angular momenta. Theoretical models can explain these correlations in terms of excitations of rotational modes of the fragments during the fission process~\cite{Moretto1980, Randrup2021, Randrup2014}. With increasing excitation energy of the fissioning system, these modes are excited more and give rise to an increase in the fragment angular momenta. 

An enhancement in the emission of isotropic discrete $\gamma$ rays was observed at higher $\gamma$-ray energies. These emissions are predominantly from stretched electric quadrupole transitions from heavy fragments with masses close to the shell closure of \ce{^{132}Sn}. The results of our analysis indicates that the angular momenta of these fragments are not strongly polarized in the plane perpendicular to the fission axis. These results can be explained by the nearly spherical shape of the fragments in this region, which makes it harder to generate angular momentum and align it. 

In addition to providing further evidence for positive correlations between the energy and angular momentum of a fragment, the results also indicate that statistical $\gamma$-ray emission is not dominantly stretched radiation. The impact of transitions where the angular momentum of the initial and final state is equivalent is strongly affected  by the energy-dependent level densities. It will be interesting to investigate the magnitude of the mixing theoretically and compare model calculations with the experimental results shown here. Understanding the magnitude of angular momentum carried by statistical $\gamma$ rays is an essential step toward in the determination of the fission fragment initial angular momenta. 

Having experimentally determined the alignment of neutrons and $\gamma$ rays, the next step will be to refine the current theoretical models and generate predictions to compare to experiment. Along with several other important observables, 
these correlations will provide significant help for the refinement of the modeling of angular momentum in fission. The improved models will, in turn, shed light on the fundamental dynamics of the fission process~\cite{Marevic2021} as well as provide predictions of $n$-$\gamma$ emission where experimental data is lacking.\\

\acknowledgments~\\[-4ex]
S.M. thanks the Chi-Nu experimental group at LANSCE-LANL and M.J. Marcath for sharing the experimental data used in this analysis. This work was in part supported by the Office of Defense Nuclear Nonproliferation Research \& Development (DNN R\&D), National Nuclear Security Administration, US Department of Energy. This work was funded in-part by the Consortium for Monitoring, Technology, and Verification under Department of Energy National Nuclear Security Administration award number DE-NA0003920. The work of V.A.P.  was performed under the auspices of UT-Battelle, LLC under Contract No. DE-AC05-00OR22725 with the U.S. Department of Energy. The work of R.V. was performed under the auspices of the U.S. Department of Energy by Lawrence Livermore National Laboratory under Contract DE-AC52-07NA27344. J. R. acknowledges support from the Office of Nuclear Physics in the U.S. Department of Energy under Contract DE-AC02-05CH11231.

\bibliography{mybib}



\end{document}